\documentclass[11pt]{article}
\usepackage{amsmath,amssymb,amsthm,mathrsfs,footmisc}
\usepackage{color}
\usepackage{graphicx}
\usepackage{hyperref}
\usepackage{float}
\usepackage{cite}
\usepackage[utf8]{inputenc}
\usepackage{tikz}
\usepackage{verbatim}
\usepackage[normalem]{ulem}

\numberwithin{equation}{section}

\newcommand{\cL}{\mathcal{L}}
\newcommand{\cO}{\mathcal{O}}

\newcommand{\hatt}{\widehat}
\newcommand{\till}{\widetilde}
\newcommand{\pd}{\partial}
\newcommand{\dd}{\text{d}}
\newcommand{\df}{\text{d}}

\newcommand{\tr}{\text{tr}}

\newcommand{\bT}{\textbf{T}}

\begin{document}
	
	\thispagestyle{empty}

	\begin{center}
		{\Large\textbf{\mathversion{bold}
Geometric realization via irrelevant deformations induced by the stress-energy tensor
}
			\par}
		
		\vspace{0.5cm}
		
		{ Xi-Yang Ran$^{1\dagger}$, Feng Hao$^{1*}$ and Masatoshi Yamada$^{1\S}$\footnotetext{All authors contributed equally to this work}	}\\
			\vspace*{0.2cm}
			{\it
				$^{1}$Center for Theoretical Physics and College of Physics, Jilin University, 
				Changchun 130012, China\\
			}
			
			\vspace*{0.5cm}
			{E-mails: {\tt ${}^\dagger$ranxy23@mails.jlu.edu.cn, ${}^*$haofeng22@mails.jlu.edu.cn, ${}^\S$yamada@jlu.edu.cn.
			}}
			\vspace{1cm}

		

\par\vspace{1.5cm}
		
\textbf{Abstract} \vspace{3mm}
		
\begin{minipage}{\textwidth}

In this paper, we generalize the deformations driven by the stress-energy tensor $T$ and investigate their relation to the flow equation for the background metric at the classical level. 
For a deformation operator $\mathcal{O}$ as a polynomial function of the stress-energy tensor, we develop a formalism that relates a deformed action to a flow equation for the metric in arbitrary spacetime dimensions.
It is shown that in the $T\bar{T}$ deformation and the $\mathcal{O}(T)=\text{tr}[\textbf{T}]^m$ deformation, the flow equations for the metric allow us to directly obtain exact solutions in closed forms. 
We also demonstrate the perturbative approach to find the same results.
As several applications of the $\mathcal{O}(T)=\text{tr}[\textbf{T}]^m$ deformation, we discuss the relation between the deformations and gravitational models.
Besides, we also deform the Lagrangians for scalar field theories.

\vspace{10pt}

\end{minipage}
		
\end{center}
	
\vspace{1.5cm}
	
\newpage
\tableofcontents
\newpage

\section{Introduction}

The $T\bar{T}$ deformation \cite{Cavaglia:2016oda,Smirnov:2016lqw} is a well-known irrelevant deformation for two-dimensional Euclidean quantum field theory. It is defined through the irrelevant composite operator~\cite{Zamolodchikov:2004ce}
\begin{equation}
\label{eq: intro TT}
    \mathcal{O}(T)=-\det[\bT] =-\frac{1}{2}\epsilon_{\mu\rho}\epsilon_{\nu\sigma}T^{\mu\nu}T^{\rho\sigma},
\end{equation}
where the $T^{\mu\nu}$ is the stress-energy tensor whose matrix form is denoted by $\bT$, and $\epsilon_{\mu\nu}$ is the Levi-Civita symbol with $\epsilon_{12}=1$. The deformation by Eq.~\eqref{eq: intro TT} exhibits many remarkable properties. For instance, in the deformed theory, one can exactly compute various physical quantities such as the finite-volume spectrum~\cite{Cavaglia:2016oda, Smirnov:2016lqw}, entanglement entropy~\cite{Donnelly:2018bef, Chen:2018eqk, Murdia:2019fax, Ota:2019yfe, Banerjee:2019ewu, Jeong:2019ylz, He:2023obo}, the $S$-matrix~\cite{Castillejo:1955ed, Mussardo:1999aj, Smirnov:2016lqw, Dubovsky:2012wk, Dubovsky:2013ira}, and the partition function on a torus~\cite{Cardy:2018sdv, Datta:2018thy, Dubovsky:2018bmo, He:2020cxp, He:2019vzf}. 
Furthermore, the deformed theory may be related to various topics in theoretical physics such as holography~\cite{McGough:2016lol, Kraus:2018xrn, Cottrell:2018skz, Taylor:2018xcy, Hartman:2018tkw, Caputa:2019pam, Guica:2019nzm}, string theory~\cite{Dubovsky:2012wk,Dubovsky:2012sh,Caselle:2013dra,Baggio:2018gct,Chen:2018keo,Chakraborty:2019mdf,Tolley:2019nmm}, and quantum gravity~\cite{Dubovsky:2017cnj, Dubovsky:2018bmo, Tolley:2019nmm, Iliesiu:2020zld,Okumura:2020dzb,Ebert:2022ehb}. See, e.g, review~\cite{Jiang:2019epa} on the $T\bar{T}$ deformation.

In addition to these properties, another remarkable aspect of the $T\bar{T}$ deformation is a deep connection with spacetime geometry. 
From a viewpoint of the field-theory side, the zweibein formalism allows us to reveal that a field theory with the $T\bar{T}$ deformation term in two-dimensional spacetime is rewritten as the original theory coupled to a two-dimensional Jackiw-Teitelboim (JT)-like gravity~\cite{Okumura:2020dzb, Dubovsky:2017cnj, Dubovsky:2018bmo, Tolley:2019nmm, Ebert:2022ehb}.
Thus, in the geometric approach, the $T\bar{T}$-deformed theory can be considered as the field theory coupled to a random geometry~\cite{Cardy:2018sdv} instead of the deformation of the theory in a dynamical coordinate system~\cite{Dubovsky:2017cnj,Conti:2018tca,Cardy:2019qao,Hirano:2024eab}.

On the other hand, the extension of the notion of the $T\bar{T}$ deformation has also attracted our attention. A marginal deformation inspired by the $T\bar{T}$ deformation is called the root-$T\bar{T}$ deformation~\cite{Ferko:2022cix,Babaei-Aghbolagh:2022leo,Babaei-Aghbolagh:2022uij,Babaei-Aghbolagh:2024uqp} in which it was pointed out that the flow obtained from the root-$T\bar{T}$ deformation is closely related to the ModMax theory. 
In the holographic perspective, the generalized notion of the $T\bar{T}$ deformation provides the cutoff condition on AdS${}_{d+1}$ in the bulk coordinate~\cite{Taylor:2018xcy, Hartman:2018tkw}. 
In the deformed Lagrangian perspective, the deformation of the free scalar field theory is related to the scalar sector in the Dirac-Born-Infeld action in $d$-dimensional spacetime~\cite{Ferko:2023sps}. 
It was shown in Ref.~\cite{Bonelli:2018kik} that the deformation operator $\det[\bT]^{1/d-1}$ can be used to solve the Lagrangian in closed form.

In Refs.~\cite{Babaei-Aghbolagh:2024hti,Tsolakidis:2024wut}, the vielbein formalism for deformed field theories has been discussed.
The deformation triggered by a certain class of operators involves the change of the background metric in higher dimensions~\cite{Conti:2022egv}.
We call this fact ``geometric realization."
Furthermore, when matter fields couple to Einstein gravity, the deformation could transform Einstein gravity to another type of gravity~\cite{Morone:2024ffm}.
Because of such attractive features, it is fascinating to investigate the geometric perspective in the generalized notion of the $T\bar{T}$ deformation.

In this work, we focus mainly on the background metric deformation in a deformed action triggered by an operator. 
It is shown that imposing the dynamical equivalence of actions with respect to the matter field infers the existence of the flow equation for the metric $\bar g_{\mu\nu}$ as a function of the deformation parameter.
The flow equations could exhibit exact solutions in the cases of the $T\bar{T}$ and the deformation operator $\cO(T)=\tr[\bT]^m$. In particular, the solution for the metric in the latter deformation might be interpreted as a Weyl transformation. 
Motivated by Ref.~\cite{Morone:2024ffm}, we consider the relation between the deformed action coupled to Einstein gravity and the undeformed action coupled to $f(R)$ gravity.
Finally, we discuss how the deformation by the operator $\cO(T)=\tr[\bT]^m$ deforms the Lagrangian for a free massless boson in arbitrary spacetime dimensions and an interactive boson in two-dimensional spacetime.

This paper is organized as follows. In Sec.~\ref{sec:Flow equation for metric from polynomial deformation}, we consider that the deformation operator $\mathcal{O}$ is a polynomial function of the stress-energy tensor. In Sec.~\ref{sec:Introduction of auxiliary fields}, we derive the flow equation for the metric from the deformation by introducing auxiliary fields. In Sec.~\ref{sec:Applications of metric flow}, we give the $T\bar{T}$ deformation and $\mathcal{O}(T)=\tr[\bT]^3$ as two examples to demonstrate the method to introduce the auxiliary fields and the derivation of the flow equations for the metric.
In Sec.~\ref{sec:Solutions to flow equation for metric}, we solve the flow equation for the metric. In Sec.~\ref{sec:The flow equation for stress energy tensor}, we give the derivation of the flow equation for the stress-energy tensor to solve the flow equation for the metric.
In Sec.~\ref{sec:Exactly solution cases}, we consider the $T\bar{T}$ deformation and the $\cO(T)=\tr[\bT]^m$ deformation and find exact solutions for the metric. In Sec~.\ref{sec:Perturbative method}, we introduce the perturbative method as another procedure for solving the flow equations. 
In Sec.~\ref{sec: Application of TrT^n}, we discuss several applications of $\cO(T)=\tr[\bT]^m$.

\section{Flow equation for metric from polynomial deformation}
\label{sec:Flow equation for metric from polynomial deformation}

In this section, we aim to derive the flow equation for the metric from the deformation. To this end, we start with the deformed action such that
\begin{align}\label{def of defor}
    \frac{\pd S_\tau[\phi,\bar g_{\mu\nu}]}{\pd \tau}=\int \dd^dx\sqrt{\bar g}\mathcal{O}_\tau(\bar T_\tau),
\end{align}
where $\phi$ and $\bar g_{\mu\nu}$ denote the unspecified matter fields and a curved spacetime background metric in the Euclidean signature, respectively. 
Here a deformation operator $\mathcal{O}_\tau(\bar T_\tau)$ is a function of the stress-energy tensor defined as
\begin{align}\label{def of stress-energy tensor}
    \bar T^{\mu\nu}_{\tau}=\frac{-2}{\sqrt{\bar g}}\frac{\delta S_\tau[\phi,\bar g_{\mu\nu}]}{\delta \bar g_{\mu\nu}},
\end{align}
with a deformation parameter $\tau$ to be positive. In this work, we assume the deformation operator to be a polynomial function of the stress-energy tensor.
Note that $\phi$ and $\bar g_{\mu\nu}$ are independent of $\tau$.

For later convenience, we write Eq.~\eqref{def of defor} equivalently as
\begin{align}\label{equi def of defor}
    S_{\tau+\delta\tau}[\phi,\bar g_{\mu\nu}]=S_\tau[\phi,\bar g_{\mu\nu}]+\delta\tau\int \dd^d x\sqrt{\bar g}\mathcal{O}_\tau(\bar T_\tau),
\end{align}
which is the first order expansion for infinitesimal $\delta\tau$.

\subsection{Introduction of auxiliary fields and flow equation for metric}
\label{sec:Introduction of auxiliary fields}
Let us describe the general idea how to obtain the flow equation for the metric.
First, we introduce auxiliary fields such that Eq.~\eqref{equi def of defor} is  decomposed into the following form:
\begin{align}\label{eq:Shat}
    \hatt S_{\tau+\delta\tau}[\phi,\bar g_{\mu\nu},h_i]=S_\tau[\phi,\bar g_{\mu\nu}]+\delta\tau\int \dd^dx\sqrt{\bar g}\left(\mathcal{F}[h_i]+\bar T^{\mu\nu}_\tau\mathscr{G}_{\mu\nu}[h_i]\right),
\end{align}
where $h_i$ are auxiliary fields. 
The number of auxiliary fields depends on the power of $\bar T_\tau$ in $\mathcal{O}_\tau(\bar T_\tau)$. 
Here, $\mathcal{F}[h_i]$ and $\mathscr{G}_{\mu\nu}[h_i]$ are quadratic functions of $h_i$.\footnote{
By ``quadratic," we mean that $\mathcal{F}[h_i]$ and $\mathscr{G}_{\mu\nu}[h_i]$ are at most quadratic functions of each $h_i$. For example, in the $\mathcal{O}_\tau(T_\tau) = \operatorname{tr}[\mathbf{T}]^3$ deformation discussed in Eq.~\eqref{eq:aux in trT3}, $\mathcal{F}[h_i] = \mathcal{F}[h_{\mu\nu}, E_{\mu\nu}]$ is quadratic in $E_{\mu\nu}$ and linear in $h_{\mu\nu}$, whereas $\mathscr{G}_{\mu\nu}[h_i] = \mathscr{G}_{\mu\nu}[h_{\mu\nu}, E_{\mu\nu}]$ is quadratic in $h_{\mu\nu}$ and linear in $E_{\mu\nu}$.
}
Note that the starting action~\eqref{equi def of defor} is reproduced from the action \eqref{eq:Shat} with the on-shell values ${h^*_i}$ which are solutions to the equation of motion for $h_i$
\begin{align}\label{eq:EOM for auxiliary fields}
    \left.\frac{\delta\hatt{S}_{\tau+\delta\tau}[\phi,\bar g_{\mu\nu},h_i]}{\delta h_j}\right|_{h_j=h^*_j}=\left.\frac{\pd \mathcal{F}[h_i]}{\pd h_j}+\bar T^{\mu\nu}_\tau\frac{\pd\mathscr{G}_{\mu\nu}[h_i]}{\pd h_j}\right|_{h_j=h^*_j}=0.
\end{align}
See Appendix~\ref{appsec: Auxiliary field} for the auxiliary field method within the path integral formalism in case of $T\bar T$ deformation.
We consider the first variation for the action~\eqref{eq:Shat} with respect to the matter field,
\begin{align}
    \begin{aligned}
        \frac{\delta\hatt{S}_{\tau+\delta\tau}[\phi,\bar g_{\mu\nu},h_j]}{\delta \phi}&=\frac{\delta}{\delta\phi}\left(S_{\tau}[\phi,\bar g_{\mu\nu}]+\delta\tau\int \dd^dx\sqrt{\bar g}\bar T^{\mu\nu}_\tau\mathscr{G}_{\mu\nu}[h_i]\right)\\
        &=\frac{\delta S_{\tau}[\phi,\bar g_{\mu\nu}-2\delta\tau\mathscr{G}_{\mu\nu}[h_i]]}{\delta\phi},
    \end{aligned}
    \label{eq:variation of phi in Shat}
\end{align}
where we have used the fact that $\mathcal F[h_i]$ is independent of $\phi$ in the first line and in the second line we have employed the definition of the stress-energy tensor~\eqref{def of stress-energy tensor}. 

Finally, we impose that the equation of motion for the matter fields are equivalent between Eqs.~\eqref{def of stress-energy tensor} and \eqref{eq:Shat} with the on-shell value $h^*_i$, i.e.,
\begin{align}\label{condition 2}
    \left.\frac{\delta\hatt{S}_{\tau+\delta\tau}[\phi,\bar g_{\mu\nu},h_i]}{\delta \phi}\right|_{h_i=h^*_i}=\frac{\delta S_{\tau+\delta\tau}[\phi,\bar g_{\mu\nu}]}{\delta \phi}.
\end{align}
We should stress here that the variation of $\hatt{S}_{\tau+\delta\tau}[\phi,\bar g_{\mu\nu},h_i]$ with respect to $\phi$ is performed before substituting the on-shell value $h^*_i$ into $h_i$. 
Together with Eq.~\eqref{eq:variation of phi in Shat}, the condition \eqref{condition 2} implies
\begin{align}\label{condition 2 equivalent}
    \frac{\delta S_{\tau+\delta\tau}[\phi,\bar g_{\mu\nu}]}{\delta\phi}=\left.\frac{\delta S_{\tau}[\phi,\bar g_{\mu\nu}-2\delta\tau\mathscr{G}_{\mu\nu}[h_i]]}{\delta\phi}\right|_{h_i=h^*_i}.
\end{align}
In terms of the matter field dynamics, the infinitesimal deformed action~\eqref{equi def of defor} is equivalent to the deformation of the background metric so that $\bar g_{\mu\nu}+\delta \bar g_{\mu\nu}$ with $\delta \bar g_{\mu\nu}= -2\delta\tau\mathscr{G}_{\mu\nu}[h_i^*]$. 

Since the equation of motion for the matter field $\phi$ is given by $\frac{\delta S_{\tau+\delta\tau}[\phi, \bar{g}_{\mu\nu}]}{\delta\phi} = 0$, we interpret the equivalence of the variations of $S_{\tau+\delta\tau}[\phi, \bar{g}_{\mu\nu}]$ and $S_{\tau}[\phi, \bar{g}_{\mu\nu} - 2\delta\tau\mathscr{G}_{\mu\nu}[h_i^*]]$ with respect to $\phi$ as indicating that both yield the same dynamics for the matter field $\phi$~\cite{Conti:2022egv}. 
Consequently, the infinitesimally deformed action in Eq.~\eqref{equi def of defor} is equivalent, at the level of dynamics, to a deformation of the background metric, such that $\bar{g}_{\mu\nu} + \delta \bar g_{\mu\nu}$, with $\delta \bar g_{\mu\nu} = -2\delta\tau\mathscr{G}_{\mu\nu}[h_i^*]$.

Here the background metric $\bar g_{\mu\nu}$ is arbitrary, so that the dynamical equivalence~\eqref{condition 2 equivalent} preserves even if we replace the background metric $\bar g_{\mu\nu}$ with a $\tau$-dependent metric $g_{\mu\nu}$.
Thus we have
\begin{align}\label{eq:dynamical equivalence in g}
    \frac{\delta S_{\tau+\delta\tau}[\phi,g_{\mu\nu}]}{\delta\phi}=\frac{\delta S_{\tau}[\phi,g_{\mu\nu}+\delta\tau\frac{\dd g_{\mu\nu}}{\dd\tau}]}{\delta\phi}.
\end{align}
This relation leads to the following flow equation for metric $g_{\mu\nu}$
\begin{align}\label{flow G}
    \frac{\dd g_{\mu\nu}}{\dd\tau}=-2{\mathscr{G}}_{\mu\nu}[h^*_i].
\end{align}
For the $\tau$-dependent metric $g_{\mu\nu}$, the definition of the deformation~\eqref{equi def of defor} becomes
\begin{align}\label{eq:def of deform in g}
    S_{\tau+\delta\tau}[\phi,g_{\mu\nu}]=S_\tau[\phi,g_{\mu\nu}]+\delta\tau\int \dd^d x\sqrt{g}\mathcal{O}_\tau(T_\tau),
\end{align}
with the stress-energy tensor
\begin{align}\label{eq:stress-energy tensor in g}
    T^{\mu\nu}_{\tau}=\frac{-2}{\sqrt{g}}\frac{\delta S_\tau[\phi,g_{\mu\nu}]}{\delta g_{\mu\nu}}.
\end{align}

For specific deformation operator $\mathcal{O}_\tau(T_\tau)$, we can find the explicit form of 
${\mathscr{G}}_{\mu\nu}[h^*_i]$. 
Nonetheless, we can show that the form of $\mathscr{G}_{\mu\nu}[h^*_j]$ is given by a function of the stress-energy tensor in the general case. 
To see this fact, we write the left-hand side of Eq.~\eqref{eq:dynamical equivalence in g} as
\begin{align}\label{relation 1}
    \begin{aligned}
        &\frac{\delta S_{\tau+\delta\tau}[\phi,g_{\mu\nu}]}{\delta\phi(x)}=\frac{\delta S_\tau[\phi,g_{\mu\nu}]}{\delta\phi}+\delta\tau\frac{\delta}{\delta\phi}\int \dd^dx\sqrt{g}\mathcal{O}_\tau(T_\tau)\\
        =&\frac{\delta S_\tau[\phi,g_{\mu\nu}]}{\delta\phi(x)}+\delta\tau\left[\sqrt{g}\frac{\pd T^{\mu\nu}_\tau(x)}{\pd\phi(x)}\frac{\pd\mathcal{O}_\tau(T_\tau)}{\pd T^{\mu\nu}_\tau(x)}+\sum^n_{i=1}(-1)^i\pd_{\mu_1}\dots\pd_{\mu_i}\left(\sqrt{g}\frac{\pd T^{\mu\nu}_\tau(x)}{\pd(\pd_{\mu_1}\dots\pd_{\mu_i}\phi(x))}\frac{\pd\mathcal{O}_\tau(T_\tau)}{\pd T^{\mu\nu}_\tau(x)}\right)\right],
    \end{aligned}
\end{align}
while using Eq.~\eqref{flow G}, the right-hand side of Eq.~\eqref{eq:dynamical equivalence in g} reads
\begin{align}\label{relation 2}
    \begin{aligned}
        &\frac{\delta S_{\tau}[\phi,g_{\mu\nu}-2\delta\tau\mathscr{G}_{\mu\nu}[h_i^*]]}{\delta\phi(x)}=\frac{\delta}{\delta\phi(x)}\left(S_{\tau}[\phi,g_{\mu\nu}]+\delta\tau\int \dd^dx\sqrt{g} T^{\mu\nu}_\tau\mathscr{G}_{\mu\nu}[h_i^*]\right)\\
        =&\frac{\delta S_\tau[\phi,g_{\mu\nu}]}{\delta\phi(x)}+\delta\tau\left[\sqrt{g}\frac{\pd T^{\mu\nu}_\tau(x)}{\pd\phi(x)}{\mathscr{G}}_{\mu\nu}[h^*_i]+\sum^n_{i=1}(-1)^i\pd_{\mu_1}\dots\pd_{\mu_i}\left(\sqrt{g}\frac{\pd T^{\mu\nu}_\tau(x)}{\pd(\pd_{\mu_1}\dots\pd_{\mu_i}\phi(x))}{\mathscr{G}}_{\mu\nu}[h^*_i]\right)\right].
    \end{aligned}
\end{align}
Comparing Eqs.~\eqref{relation 1} and \eqref{relation 2}, we obtain
\begin{align}\label{G O equation}
    \mathscr{G}_{\mu\nu}[h^*_i]=\frac{\pd\mathcal{O}_\tau(T_\tau)}{\pd T^{\mu\nu}_\tau}.
\end{align}
Thus, we finally arrive at the flow equation of the metric triggered by $\mathcal{O}_\tau(T_\tau)$ such that
\begin{align}\label{metric flow -}
    \frac{\dd g_{\mu\nu}}{\dd\tau}=-2\frac{\pd\mathcal{O}_\tau(T_\tau)}{\pd T^{\mu\nu}_\tau}.
\end{align}
Equation~\eqref{metric flow -} allows us to obtain the flow equation for the metric directly from the deformation operator $\mathcal{O}_\tau(T_\tau)$ without the introduction of auxiliary fields.


To summarize, we have shown that introducing the auxiliary fields in the form of Eq.~\eqref{eq:Shat} reveals a dynamical equivalence between the deformed action and the deformed background spacetime metric. After replacing the background metric $\bar g_{\mu\nu}$ with a $\tau$-dependent one $g_{\mu\nu}$ we have
\begin{align}\label{statement S g}
        &S_{\tau+\delta\tau}[\phi,g_{\mu\nu}]\simeq S_{\tau}\left[\phi,g_{\mu\nu}+\frac{\dd g_{\mu\nu}}{\dd\tau}\delta\tau\right],
\end{align}
where symbol $\simeq$ denotes the dynamical equivalence with respect to matter fields \cite{Conti:2022egv}. 
Equation~\eqref{statement S g} indicates that the deformation of the action (the left-hand side) corresponds to the deformation of the background spacetime metric (the right-hand side).
From this fact, we can read off the flow equation for the metric so as to be Eq.~\eqref{metric flow -}.
An advantage of our formula~\eqref{metric flow -} is that one can directly obtain the flow equation of the metric from the dependence of the deformation operators on the stress-energy tensor, i.e., one does not have to obtain the explicit solutions to the equations of motion for the auxiliary fields.

In the following subsections, we demonstrate the derivation of the flow equation for the metric in several explicit deformation operators.

\subsection{Applications}\label{sec:Applications of metric flow}
In this subsection, we apply Eq.~\eqref{metric flow -} for the $T\bar{T}$ deformation in two-dimensional spacetime and the deformation by the operator $\mathcal{O}_\tau(T_\tau)=\tr[\bT_\tau]^3$ in arbitrary spacetime dimensions. 

\subsubsection{Review: $T\bar{T}$ deformation}
First, we consider the well-known $T\bar{T}$ deformation which is triggered by the deformation operator $\mathcal{O}_\tau(T_\tau)=\frac{1}{2}(g_{\mu\nu} g_{\rho\sigma}-g_{\mu\rho} g_{\nu\sigma})T^{\mu\nu}_\tau T^{\rho\sigma}_\tau$. The deformed action is given by
\begin{align}\label{TT action}
    \begin{aligned}
        S_{\tau+\delta\tau}[\phi,g_{\mu\nu}]=S_{\tau}[\phi,g_{\mu\nu}]+\frac{\delta\tau}{2}\int \dd^2x\sqrt{g}(g_{\mu\nu} g_{\rho\sigma}-g_{\mu\rho} g_{\nu\sigma})T^{\mu\nu}_\tau T^{\rho\sigma}_\tau.
    \end{aligned}
\end{align}
Using Eq.~\eqref{metric flow -}, we obtain the flow equation of the metric as
\begin{align}
\label{eq:TTbar flow equation of metric}
    \frac{\dd g_{\mu\nu}}{\dd\tau}=-2(g_{\mu\nu} g_{\rho\sigma}-g_{\mu\rho} g_{\nu\sigma})T^{\rho\sigma}_\tau.
\end{align}
This result was shown in Ref.~\cite{Conti:2022egv} by using the auxiliary field method.

Let us here show the explicit derivation of Eq.~\eqref{eq:TTbar flow equation of metric} by the auxiliary field method for the $T\bar{T}$ deformation. 
We start with the definition of the deformation Eq.~\eqref{eq:stress-energy tensor in g}.
Since the deformation operator involves the quadratic form of the stress-energy tensor, an auxiliary field $h_{\mu\nu}$ should be introduced such that the deformed action is written as\footnote{In Appendix~\ref{appsec: Auxiliary field}, we exhibit the explicit way of the introduction of the auxiliary field for the $T\bar{T}$-deformed action \eqref{TT action} within the path integral formalism.}
\begin{align}\label{TT auxiliary action}
    \hatt{S}_{\tau+\delta\tau}[\phi,g_{\mu\nu},h_{\mu\nu}]=S_{\tau}[\phi,g_{\mu\nu}]+\delta\tau\int \dd^2x\sqrt{g}\left[c(g^{\mu\nu} g^{\rho\sigma}- g^{\mu\rho} g^{\nu\sigma})h_{\mu\nu}h_{\rho\sigma}-\frac{1}{2}T^{\mu\nu}_\tau h_{\mu\nu}\right],
\end{align}
where $c$ is an undetermined constant. 
Solving the equation of motion for $h_{\mu\nu}$, i.e., 
\begin{align}
\left.\frac{\delta\hatt S_{\tau+\delta\tau}[\phi,g_{\mu\nu},h_{\mu\nu}]}{\delta h_{\mu\nu}}\right|_{h_{\mu\nu}=h^*_{\mu\nu}}=0,
\end{align}
we obtain the on-shell value
\begin{align}\label{TT h on-shell}
    h^*_{\mu\nu}=\frac{1}{4c}(g_{\mu\nu} g_{\rho\sigma}- g_{\mu\rho} g_{\nu\sigma})T^{\rho\sigma}_\tau.
\end{align}

Next, from the equivalence between Eqs.~\eqref{TT action} and \eqref{TT auxiliary action} in terms of the equation of motion for $\phi$ [see Eq.~\eqref{condition 2}], we have
\begin{align}\label{TT first variation equivalence}
    \frac{\delta S_{\tau+\delta\tau}[\phi,g_{\mu\nu}]}{\delta\phi}= \left.\frac{\delta}{\delta\phi}\left(S_\tau[\phi,g_{\mu\nu}]-\frac{1}{2}\delta\tau\int \dd^2x\sqrt{g}T^{\mu\nu}_\tau h_{\mu\nu}\right)\right|_{h_{\mu\nu}=h^*_{\mu\nu}}.
\end{align}
Note here that the on-shell value has to be taken after the functional derivative with respect to $\phi$. Comparing the coefficients, one can find $c=-\frac{1}{8}$. Then, the flow equation for the metric reads
\begin{align}
    \frac{\dd g_{\mu\nu}}{\dd\tau}=h^*_{\mu\nu}=-2(g_{\mu\nu} g_{\rho\sigma}-g_{\mu\rho} g_{\nu\sigma})T^{\rho\sigma}_\tau,
\end{align}
This agrees with Eq.\eqref{eq:TTbar flow equation of metric} obtained from Eq.~\eqref{metric flow -}.

\subsubsection{$\mathcal{O}_\tau(T_\tau)=\tr[\bT_\tau]^3$ case}

Next, let us consider the deformation triggered by $\cO_\tau(T_\tau)=\tr[\bT_\tau]^3$ and derive the flow equation for the metric. In such a case, the deformed action is given by
\begin{align}
    S_{\tau+\delta\tau}[\phi,g_{\mu\nu}]=S_\tau[\phi,g_{\mu\nu}]+\delta\tau\int \dd^dx\sqrt{g}\tr[\bT_\tau]^3.
\end{align}
Here, we do not specify spacetime dimension $d$ in which the mass dimension of the deformation parameter is not 2.
The use of Eq.~\eqref{metric flow -} yields
\begin{align}
\label{eq:TTT deformation flow equation}
    \frac{\dd g_{\mu\nu}}{\dd\tau}=-6\tr[\bT_\tau]^2 g_{\mu\nu}.
\end{align}

We briefly introduce the auxiliary field method for the present case to demonstrate that the same flow equation is obtained.
Two auxiliary fields $h_{\mu\nu}$ and $E_{\mu\nu}$ need to be introduced due to the cubic form of the stress-energy tensor. 
Thus, the deformed action with the auxiliary fields takes the form
\begin{align}
    \begin{aligned}
       & \hatt{S}_{\tau+\delta\tau}[\phi,g_{\mu\nu},h_{\mu\nu},E_{\mu\nu}]=S_\tau[\phi,g_{\mu\nu}]\\
        &+\delta\tau\int \dd^dx\sqrt{g}\left(a_1 g^{\gamma\delta} g^{\rho\sigma} g_{\mu\nu}h_{\gamma\delta}h_{\rho\sigma}T_\tau^{\mu\nu}+a_2 g^{\mu\nu} g^{\rho\sigma} g^{\gamma\delta}h_{\mu\nu}E_{\rho\sigma}E_{\gamma\delta}+ g^{\gamma\delta} g^{\rho\sigma} g_{\mu\nu}h_{\gamma\delta}E_{\rho\sigma}T_\tau^{\mu\nu}\right),
    \end{aligned}
\end{align}
with undetermined constants $a_1$ and $a_2$.
This implies that the functions introduced in Eq.~\eqref{eq:Shat} read
\begin{align}\label{eq:aux in trT3}
    \begin{aligned}
        &\mathcal{F}[h_{\mu\nu},E_{\mu\nu}]=a_2 g^{\mu\nu} g^{\rho\sigma} g^{\gamma\delta}h_{\mu\nu}E_{\rho\sigma}E_{\gamma\delta},\\
        &\mathscr{G}_{\mu\nu}[h_{\mu\nu},E_{\mu\nu}]=\left(a_1 g^{\gamma\delta} g^{\rho\sigma}h_{\gamma\delta}h_{\rho\sigma}+ g^{\gamma\delta} g^{\rho\sigma}h_{\gamma\delta}E_{\rho\sigma}\right) g_{\mu\nu}.
    \end{aligned}
\end{align}
Following the process in Sec.~\ref{sec:Introduction of auxiliary fields}, the on-shell values of $h_{\mu\nu}$ and $E_{\mu\nu}$ are
\begin{align}\label{eq:T3 on-shell}
    \tr[E^*]=-\frac{1}{2a_2}\tr[\bT_\tau],\qquad \tr[h^*]=\frac{1}{8a_1a_2}\tr[\bT_\tau].
\end{align}
Substituting Eq.~\eqref{eq:T3 on-shell} to Eqs.~\eqref{eq:dynamical equivalence in g} and \eqref{flow G}, we obtain $a_1a_2^2=-\frac{1}{64}$ and then the flow equation for the metric reads
\begin{align}
    \frac{\dd g_{\mu\nu}}{\dd\tau}=-6\tr[\bT_\tau]^2g_{\mu\nu}.
\end{align}
This is the same equation as \eqref{eq:TTT deformation flow equation}.

\subsubsection{General deformation of order $N$}

Our formula~\eqref{metric flow -} is applicable to a more general class of deformation operators.
A general order-$N$ deformation is given by
\begin{align}\label{eq:general O}
    \mathcal{O}_\tau(T_\tau)=\sum_{\{a_i,b_i\}}C_{\{a_i,b_i\}}\tr[\bT_\tau^{a_1}]^{b_1}\dots\tr[\bT_\tau^{a_n}]^{b_n},\qquad \sum_{\{a_i,b_i\}}a_ib_i=N,
\end{align}
where $C_{\{a_i,b_i\}}$ are arbitrary constants and $a_i\neq a_j$ for $i\neq j$. 
We demonstrate the convenience of Eq.~\eqref{metric flow -} by applying it to a class of deformation operators characterized by Eq.~\eqref{eq:general O}.

A simple example is the deformation triggered by
\begin{align}\label{eq:T3 defor 2nd}
\mathcal{O}_\tau(T_\tau)=\tr[\bT_\tau^3]
=g^{\mu\nu}(T_{\mu\rho,\tau}g^{\rho\sigma}T_{\sigma\gamma,\tau}g^{\gamma\delta}T_{\delta\nu,\tau}),
\end{align}
which corresponds to $a_1=3$, $b_1=1$, $C_{a_1,b_1}=1$, and $C_{\{a_i,b_i\}}=a_i=b_i=0$ ($i>1$) for Eq.~\eqref{eq:general O}. 
Utilizing Eq.~\eqref{metric flow -}, we obtain the flow equation for the metric as
\begin{align}\label{eq:T3 flow eq 2nd}
    \frac{\dd g_{\mu\nu}}{\dd\tau}=-6T_{\mu\rho,\tau} g^{\rho\sigma}T_{\sigma\nu,\tau}.
\end{align}

Another example is the deformation operator
\begin{align}
\mathcal{O}_\tau(T_\tau)=\tr[\bT_\tau^2]\tr[\bT_\tau]^2,
\end{align}
which corresponds to $a_1=2$, $b_1=1$, $a_2=1$, $b_2=2$, $C_{\{a_1,b_1\}}=C_{\{a_2,b_2\}}=1$ and $C_{\{a_i,b_i\}}=a_i=b_i=0$ for $i>2$.
The use of Eq.~\eqref{metric flow -} gives
\begin{align}\label{eq:flow eq T4}
    \frac{\dd g_{\mu\nu}}{\dd\tau}=-4\tr[\bT_\tau]^2T_{\mu\nu,\tau}-4\tr[\bT_\tau^2]\tr[\bT_\tau]g_{\mu\nu}.
\end{align}
As demonstrated, the flow equation can be derived straightforwardly.
The auxiliary field method for above two cases yields the same flow equations. 
In addition, different deformation operators require introducing various numbers of auxiliary fields. In the formalism given by Eq.~\eqref{eq:general O}, the number of auxiliary fields needed for such an operator is 
$$
\sum_{\{a_j,b_j\}} j \lfloor \log N \rfloor,
$$
where \( \lfloor \cdot \rfloor \) denotes the floor function. We illustrate this with two examples in Appendix~\ref{App:T3 and T4}.

\section{Solutions to flow equation for metric}\label{sec:Solutions to flow equation for metric}
In this section, we aim to give several solutions to the flow equation for the metric. To this end, we need to have the flow equation for the stress-energy tensor.
In this process, it is convenient to rewrite the flow equation for the metric by the transformations $\delta\tau\to -\delta\tau$ and $\tau\to \tau+\delta\tau$ for which we have
\begin{align}\label{statement S g_1}
        &S_{\tau}[\phi,g_{\mu\nu}]\simeq S_{\tau+\delta\tau}\left[\phi,g_{\mu\nu}+\frac{\dd g_{\mu\nu}}{\dd\tau}\delta\tau\right],
\end{align}
and
\begin{align} \label{metric flow +}
\frac{\dd g_{\mu\nu}}{\dd\tau}=2\frac{\pd\mathcal{O}_\tau[T_\tau]}{\pd T^{\mu\nu}_\tau}.
\end{align}
In this form, Eq.~\eqref{statement S g_1} implies that the deformation of the action and of the metric does not change the original action if the metric obeys the flow equation~\eqref{metric flow +}.

In Sec.~\ref{sec:The flow equation for stress energy tensor}, we derive the flow equation for the stress-energy tensor utilizing Eq.~\eqref{metric flow +}. 
In Sec.~\ref{sec:Exactly solution cases}, we consider the deformations by the $T\bar T$ case and the $\cO_\tau(T_\tau)=\tr[\bT_\tau]^m$ operator. In these cases, the flow equations can be solved exactly. It turns out that some general properties of the deformation driven by the stress-energy tensor preserve even if the deformation operator is different. 
In Sec~\ref{sec:Perturbative method}, we deal with the deformation by $\cO_\tau=a_m\tr[\bT]^m$ to demonstrate the perturbative method algorithm.
In this section, we refrain from solving the flow equations \eqref{eq:T3 flow eq 2nd} and \eqref{eq:flow eq T4} due to the difficulty in obtaining their exact solutions.

\subsection{The flow equation for stress-energy tensor}
\label{sec:The flow equation for stress energy tensor}

The discussions in Sec.~\ref{sec:Flow equation for metric from polynomial deformation} focus on how to obtain the flow equation for the metric from the deformation. The deformation operator $\cO_\tau(T_\tau)$ contains $T^{\mu\nu}_\tau$, so that we also need to have the flow equation for the stress-energy tensor to solve Eq.~\eqref{metric flow +}. The flow equation for the stress-energy tensor is given by
\begin{align}\label{stress energy tensor flow equation}
    \frac{\dd T^{\mu\nu}_\tau}{\dd\tau}=\left(g^{\mu\nu}T^{\rho\sigma}_\tau-g^{\rho\sigma}T^{\mu\nu}_\tau\right)\frac{\pd\cO_\tau(T_\tau)}{\pd T^{\rho\sigma}_\tau}-g^{\mu\nu}\mathcal{O}_\tau(T_\tau)-2\frac{\partial \mathcal{O}_\tau(T_\tau)}{\pd g_{\mu\nu}}.
\end{align}
Note here that $g_{\mu\nu}$ in Eq.~\eqref{stress energy tensor flow equation} depends on the deformation parameter $\tau$.

Let us now show the derivation of this flow equation.
Our starting point is the deformed action in the right-hand side of Eq.~\eqref{statement S g_1}, i.e.,
\begin{align}
S_{\tau+\delta\tau}\left[\phi,g_{\mu\nu}+\frac{\dd g_{\mu\nu}}{\dd\tau}\delta\tau\right] \equiv S_{\tau'}\left[\phi, g_{\mu\nu}(\tau')\right],
\end{align}
where $\tau'=\tau+\delta\tau$. Thus, both the action $S_{\tau}[\phi,g_{\mu\nu}]$ and the metric $g_{\mu\nu}$ depend on the deformation parameter $\tau'$. Hereafter, we omit the prime on the deformation parameter.
Following Ref.~\cite{Guica:2019nzm}, we suppose that the functional derivative $\frac{\delta}{\delta g_{\mu\nu}}$ commutes with $\frac{\pd}{\pd \tau}$ when those act on the deformed action at any value of $\tau$. Hence, we impose\footnote{We assume that the action is always a smooth function of the deformation parameter $\tau$ and the background metric $g_\tau$.
In this case, $\frac{\delta}{\delta g_{\mu\nu}}$ and $\frac{\pd}{\pd\tau}$ can be interchanged.}
\begin{align}\label{eq:commute action}
    \frac{\delta}{\delta g_{\mu\nu}}\frac{\pd}{\pd\tau}S_{\tau}[\phi,g_{\mu\nu}]=\frac{\pd}{\pd \tau}\frac{\delta}{\delta g_{\mu\nu}}S_{\tau}[\phi,g_{\mu\nu}].
\end{align}
From Eq.~\eqref{def of stress-energy tensor}, the right-hand side of  Eq.~\eqref{eq:commute action} is
\begin{align}\label{eq:R.H.S of commute}
    \begin{aligned}
        \frac{\pd}{\pd \tau}\frac{\delta}{\delta g_{\mu\nu}}S_{\tau}[\phi,g_{\mu\nu}]=-\frac{\sqrt{g}}{2}\frac{\pd T^{\mu\nu}_\tau}{\pd\tau}.
    \end{aligned}
\end{align}
Note here that we apply the partial derivative $\frac{\partial}{\partial \tau}$ to any quantity while keeping $g_{\mu\nu}$ fixed.
The left-hand side of Eq.\eqref{eq:commute action} reads
\begin{align}\label{eq:L.H.S of commute}
    \begin{aligned}
        &\frac{\delta}{\delta g_{\mu\nu}}\frac{\pd}{\pd\tau}S_{\tau}[\phi,g_{\mu\nu}]=\frac{1}{2}\sqrt{g}\cO_\tau(T_\tau) g^{\mu\nu}+\sqrt{g}\left(\frac{\pd\cO_\tau(T_\tau)}{\pd g_{\mu\nu}}+\frac{\pd\cO_\tau(T_\tau)}{\pd T^{\rho\sigma}_\tau}\frac{\pd T^{\rho\sigma}_\tau}{\pd g_{\mu\nu}}\right)\\
        &=\frac{1}{2}\sqrt{g}\cO_\tau(T_\tau) g^{\mu\nu}+\sqrt{g}\frac{\pd\cO_\tau(T_\tau)}{\pd g_{\mu\nu}}+\frac{1}{2}\sqrt{g}\left(T^{\mu\nu}_\tau g^{\rho\sigma}-g^{\mu\nu}T^{\rho\sigma}_\tau\right)\frac{\pd\cO_\tau(T_\tau)}{\pd T^{\rho\sigma}_\tau}+\frac{1}{2}\sqrt{g}\frac{\dd g_{\rho\sigma}}{\dd\tau}\frac{\pd T^{\mu\nu}_\tau}{\pd g_{\rho\sigma}}.
    \end{aligned}
\end{align}
Here, we have used Eq.~\eqref{metric flow +} and the identity
\begin{align}
\label{eq: derivative T with g}
    \frac{\pd T^{\rho\sigma}_\tau}{\pd g_{\mu\nu}}=\frac{1}{2}g^{\rho\sigma}T^{\mu\nu}_\tau-\frac{1}{2}g^{\mu\nu}T^{\rho\sigma}_\tau+\frac{\pd T^{\mu\nu}_\tau}{\pd g_{\rho\sigma}}.
\end{align}
See Appendix~\ref{appsec: detail of derivations} for its explicit derivation.
Because the total derivative of $T^{\mu\nu}_\tau$ is given by
\begin{align}\label{eq:total derivative of stress energy tensorr}
    \frac{\dd T^{\mu\nu}_\tau}{\dd\tau}=\frac{\pd T^{\mu\nu}_\tau}{\pd \tau}+\frac{\dd g_{\rho\sigma}}{\dd\tau}\frac{\pd T^{\mu\nu}_\tau}{\pd g_{\rho\sigma}},
\end{align}
we find Eq.~\eqref{stress energy tensor flow equation} by combining Eqs.~\eqref{eq:R.H.S of commute} and \eqref{eq:L.H.S of commute}. 

In summary, we obtain the flow equations corresponding to the deformation operator $\cO_\tau(T_\tau)$ as
\begin{align}\label{eq:flow equation for g and T}
    \left\{\begin{aligned}
        \frac{\dd g_{\mu\nu}}{\dd\tau}=&2\frac{\pd\mathcal{O}_\tau(T_\tau)}{\pd T^{\mu\nu}_\tau},\\
        \frac{\dd T^{\mu\nu}_\tau}{\dd\tau}=&\left(g^{\mu\nu}T^{\rho\sigma}_\tau-g^{\rho\sigma}T^{\mu\nu}_\tau\right)\frac{\pd\cO_\tau(T_\tau)}{\pd T^{\rho\sigma}_\tau}-g^{\mu\nu}\mathcal{O}_\tau(T_\tau)-2\frac{\partial \mathcal{O}_\tau(T_\tau)}{\pd g_{\mu\nu}}.
    \end{aligned}\right.
\end{align}
We solve both Eqs.~\eqref{metric flow +} and \eqref{stress energy tensor flow equation} in the next subsection.

\subsection{Exact solutions}
\label{sec:Exactly solution cases}
As we demonstrate in this subsection, the flow equation~\eqref{eq:flow equation for g and T} gives simple solutions in the $T\bar{T}$ deformation~\cite{Guica:2019nzm, Conti:2022egv} and the $\cO_\tau(T_\tau)=\tr[\bT_\tau]^m$ deformation where $m$ is a positive integer.

\subsubsection{$T\bar{T}$ deformation case}\label{sec:TT deformation case}

For the $T\bar{T}$ deformation, i.e., $\cO_\tau(T_\tau) =\det[\bT_\tau] =\frac{1}{2}\tr[\bT_\tau]^2-\frac{1}{2}\tr[\bT_\tau^2]$, the corresponding flow equations from Eq.~\eqref{eq:flow equation for g and T} read~\cite{Guica:2019nzm, Conti:2022egv}
\begin{align}\label{TT system}
    \left\{\begin{aligned}
        &\frac{\dd g_{\mu\nu}}{\dd\tau}=2\tr[\bT_\tau]g_{\mu\nu}-2T_{\mu\nu,\tau},\\[1ex]
        &\frac{\dd T_{\mu\nu,\tau}}{\dd\tau}=-2T^2_{\mu\nu,\tau}+\tr[\bT_\tau]T_{\mu\nu,\tau}+\left(\frac{1}{2}\tr[\bT_\tau]^2-\frac{1}{2}\tr[\bT^2_\tau]\right)g_{\mu\nu}.
    \end{aligned}\right.
\end{align}
Here, we have abbreviated the products of any tensor $A_{\mu\nu}$ as
\begin{equation}
    A^k_{\mu\nu}\equiv A_{\mu a_1}g^{a_1a_2}A_{a_2a_3}\dots A_{a_{2k-4}a_{2k-3}}g^{a_{2k-3}a_{2k-2}}A_{a_{2k-2}\nu}.
\end{equation}
Using Eq.~\eqref{TT system}, we can see that
\begin{align}\label{TT system T hat}
        \frac{\dd^3g_{\mu\nu}}{\dd\tau^3}=\frac{\dd^2\hatt T_{\mu\nu,\tau}}{\dd\tau^2}=0,
\end{align}
where we have defined $\hatt T_{\mu\nu,\tau}=2\tr[\bT_\tau]g_{\mu\nu}-2T_{\mu\nu,\tau}$ and used the relation:\footnote{This equation can be proven in a $2\times 2$ matrix form by leveraging the symmetry properties of the tensors $g_{\mu\nu}$ and $T_{\mu\nu}$.}
\begin{align}
\tr[\bT^2_\tau]g_{\mu\nu}=2T^2_{\mu\nu,\tau}+\tr[\bT_\tau]^2g_{\mu\nu}-2\tr[\bT_\tau]T_{\mu\nu,\tau}.
\end{align}
Equation~\eqref{TT system T hat} means that if we expand $g_{\mu\nu}$ and $\hatt T_{\mu\nu,\tau}$ as
\begin{align}\label{eq:Taylor expansion}
    \left\{\begin{aligned}
        g_{\mu\nu}|_{\tau+\Delta\tau}=&g_{\mu\nu}|_{\tau}+\left.\frac{\dd g_{\mu\nu}}{\dd\tau}\right|_{\tau}\Delta\tau+\frac{1}{2}\left.\frac{\dd^2g_{\mu\nu}}{\dd\tau^2}\right|_{\tau}\Delta\tau^2+\dots,\\
        \hatt T_{\mu\nu,\tau+\Delta\tau}=&\hatt T_{\mu\nu,\tau}+\frac{\dd\hatt T_{\mu\nu,\tau}}{\dd\tau}\Delta\tau+\frac{1}{2}\frac{\dd^2\hatt T_{\mu\nu,\tau}}{\dd\tau^2}\Delta\tau^2+\dots,
    \end{aligned}\right.
\end{align}
these series stop at the quadratic order and then the solutions are easily found to be
\begin{align} \label{eq:TTbar flow solution}
    \left\{\begin{aligned}
        &g_{\mu\nu}|_{\tau+\Delta\tau}=g_{\mu\nu}|_{\tau}+\hatt T_{\mu\nu,\tau}\Delta\tau+\frac{1}{4}\hatt T^2_{\mu\nu,\tau}\Delta\tau^2,\\
        &\hatt T_{\mu\nu,\tau+\Delta\tau}=\hatt T_{\mu\nu,\tau}+\frac{1}{2}\hatt T^2_{\mu\nu,\tau}\Delta\tau.
    \end{aligned}\right.
\end{align}
These results are consistent with Ref.~\cite{Guica:2019nzm}. 
In addition, we notice that
\begin{align}\label{sqrt metric}
    &\frac{\dd^3\sqrt{g}}{\dd\tau^3}=2\frac{\dd}{\dd\tau}\Big(\sqrt{g}\cO_\tau(T_\tau)\Big)=0,
\end{align}
where we have used the fact $\frac{\dd\det[g]}{\dd\tau}=\det[g]g^{\mu\nu}\frac{\dd g_{\mu\nu}}{\dd\tau}$.
For the explicit derivation of Eq.~\eqref{sqrt metric}, see Appendix~\ref{appsec: detail of derivations}.
The fact in Eq.~\eqref{sqrt metric} implies that the combination $\sqrt{g}\cO_\tau(T_\tau)$ is invariant under the $T\bar{T}$ deformation.
Thus, we get a closed form of $\sqrt{g}$ as
\begin{align}    \sqrt{g}|_{\tau+\Delta\tau}=\sqrt{g}|_{\tau}\left(1+\tr[\bT_\tau]\Delta\tau+\cO_{\tau}(T_\tau)\Delta\tau^2\right).
\end{align}

\subsubsection{$\cO_\tau(T_\tau)=\tr[\bT_\tau]^m$ case}\label{sec:Exactly (TrT^M) case}
Next, we consider another case of the $\cO_\tau(T_\tau)=a_m\tr[\bT_\tau]^m$ deformation in general $d$ spacetime dimensions. Here, $a_m$ is a coupling constant for a fixed $m$. From Eq.~\eqref{eq:flow equation for g and T}, we obtain
\begin{align}\label{trT m}
    \left\{\begin{aligned}
        &\frac{\dd g_{\mu\nu}}{\dd\tau}=2m\cdot a_m\tr[\bT_\tau]^{m-1}g_{\mu\nu},\\
        &\frac{\dd T^{\mu\nu}_\tau}{\dd\tau}=-ma_m(d+2)\tr[\bT_\tau]^{m-1}T^{\mu\nu}_\tau+(m-1)a_m\tr[\bT_\tau]^mg^{\mu\nu},
    \end{aligned}\right.
\end{align}
from which the flow equation for $\tr[\bT_\tau]$ becomes
\begin{align}\label{trT tau}
    \begin{aligned}
        \frac{\dd\tr[\bT_\tau]}{\dd\tau}
        =-da_m\tr[\bT_\tau]^m.
    \end{aligned}
\end{align}

For the initial value of deformation parameter $\tau$ at $\tau=0$, the solution to the flow equation~\eqref{trT tau} is
\begin{align}\label{eq:closed form for trT}
    \tr[\bT_\tau]^{1-m}=(m-1)da_m\tau+\tr[\bT_0]^{1-m},
\end{align}
where $\tr[\bT_0]$ stands for the value of $\tr[\bT_\tau]$ at $\tau=0$. Substituting this equation into Eq.~\eqref{trT m}, the flow equation for $g_{\mu\nu}$ is written in the form of
\begin{align}
    \frac{\dd g_{\mu\nu}}{\dd\tau}=\frac{2ma_m}{(m-1)da_m\tau+\tr[\bT_0]^{1-m}}g_{\mu\nu}.
\end{align}
Its solution reads
\begin{align}\label{trTm g closed form}
    g_{\mu\nu}|_{\tau}=\left(1+\frac{(m-1)d}{2m}\cdot 2m a_m\tr[\bT_0]^{m-1}\tau\right)^{\frac{2m}{d(m-1)}}g_{\mu\nu}|_{\tau=0},
\end{align}
We note that the effect of the deformation operator $\cO_\tau(T_\tau)=a_m\tr[\bT_\tau]^m$ on the metric might be interpreted as a Weyl transformation with a factor $\Omega(\tau)=\left(1+(m-1)da_m\tr[\bT_0]^{m-1}\tau\right)^{\frac{m}{d(m-1)}}$.

Now, inspired by Eq.~\eqref{sqrt metric}, let us explore whether or not a similar aspect is given in the case of the deformation by $\cO_\tau(T_\tau)=\tr[\bT_\tau]^m$. 
To do so, from Eq.~\eqref{eq:flow equation for g and T} for $\cO_\tau(T_\tau)$, we have
\begin{align}\label{TrT}
    \frac{\dd\tr[\bT_\tau]}{\dd\tau}=-d\cO_\tau(T_\tau)+\left(dT^{\mu\nu}_\tau-\tr[\bT_\tau]g^{\mu\nu}\right)\frac{\pd\cO_\tau(T_\tau)}{\pd T^{\mu\nu}_\tau},
\end{align}
where we have used the identity 
\begin{align}
    \frac{\pd \cO_\tau(T_\tau)}{\pd T^{\mu\nu}_\tau}T^{\mu\nu}_\tau-\frac{\pd \cO_\tau(T_\tau)}{\pd g_{\mu\nu}}g_{\mu\nu}=0.
\end{align}
The derivation of Eq.~\eqref{TrT} is presented in Appendix~\ref{appsec: detail of derivations}.
Next, the derivative of $\sqrt{g}\cO_\tau(T_\tau)$ with respect to $\tau$ yields 
\begin{align}\label{eq:sqrt g O constant}
        \frac{\dd(\sqrt{g}\cO_\tau(T_\tau))}{\dd\tau}=\frac{1}{2}\sqrt{g}g^{\mu\nu}\frac{\dd g_{\mu\nu}}{\dd\tau}\cO_\tau(T_\tau)+\sqrt{g}\frac{\dd\cO_\tau(T_\tau)}{\dd\tau}=0.
\end{align}
Finally, we arrive at the relation $\sqrt{g}\cO_\tau(T_\tau)=\sqrt{g}\cO_\tau(T_\tau)|_{\tau=0}$.
Thus, the combination $\sqrt{g}\cO_\tau(T_\tau)$ is invariant under the deformation.
Indeed, in the derivation above, we have not specified the form of the deformation operator $\mathcal O_\tau$. Hence, we find that the same fact still holds not only the case of $\cO_\tau(T_\tau)=\tr[\bT_\tau]^m$ but also any deformation operators. Note that using Eqs.~\eqref{TrT} and \eqref{eq:flow equation for g and T}, we obtain
\begin{align}
    \begin{aligned}
        \frac{\dd(\sqrt{g}\tr[\bT_\tau])}{\dd\tau}=d(m-1)\sqrt{g}\cO_\tau(T_\tau),
    \end{aligned}
\end{align}
from which the following relation is derived
\begin{align}
        \sqrt{g}\tr[\bT_\tau]=d(m-1)\sqrt{g}\cO_\tau(T_\tau)|_{\tau=0}\cdot\tau+\sqrt{g}\tr[\bT_\tau]|_{\tau=0}.
\end{align}

\subsection{Perturbative method}\label{sec:Perturbative method}
As shown in the previous subsection, we have obtained the exact solutions to the flow equation for the metric in the cases of $\cO_\tau(T_\tau)=T\bar{T}$ and $\tr[\bT_\tau]^m$.
In most cases, however, we cannot integrate Eq.~\eqref{eq:flow equation for g and T} to obtain the closed flow for deformed metric $g_{\mu\nu}$ as in Sec.~\ref{sec:Exactly (TrT^M) case}. 
In such situations, we may be able to solve Eq.~\eqref{eq:flow equation for g and T} in the perturbative method based on the expansion~\eqref{eq:Taylor expansion}.
In this subsection, we demonstrate the perturbative approach for solving $\cO_\tau(T_\tau)=a_m\tr[\bT_\tau]^m$ again as an example and show that the same result as Eq.~\eqref{trTm g closed form} is actually reproduced. 

For simplicity, we set $a_m=1$ so that $\cO_\tau=\tr[\bT_\tau]^m$. Then, Eq.~\eqref{trT m} becomes
\begin{align}\label{eq:tr[T^m] flow}
    \left\{\begin{aligned}
        &\frac{\dd g_{\mu\nu}}{\dd\tau}=\till T_{\mu\nu,\tau},\\
        &\frac{\dd\till{T}_{\mu\nu,\tau}}{\dd\tau}=\till{T}^2_{\mu\nu,\tau}+\alpha_\tau\till{T}_{\mu\nu,\tau}+\beta_\tau g_{\mu\nu},
    \end{aligned}\right.
\end{align}
where $\till T_{\mu\nu,\tau}=2m\tr[\bT_\tau]^{m-1}g_{\mu\nu}$. 
Here, the coefficients $\alpha_\tau$ and $\beta_\tau$ have been defined as
\begin{align}
    \begin{aligned}
        \alpha_\tau=-(m^2-m)d\tr[\bT_\tau]^{m-1},\quad\quad\beta_\tau=2m(m-1)^2d\tr[\bT_\tau]^{2m-2},
    \end{aligned}
\end{align}
which obey the flow equations from Eq.~\eqref{trT m} as
\begin{align}
    \left\{\begin{aligned}
        &\frac{\dd\alpha_\tau}{\dd\tau}=\frac{1}{m}\alpha_\tau^2,\\
        &\frac{\dd\beta_\tau}{\dd\tau}=\frac{2}{m}\alpha_\tau\beta_\tau.
    \end{aligned}\right.
\end{align}

Let us expand the metric $g_{\mu\nu}$ in terms of the deformation parameter as
\begin{align}
    \label{eq:g Taylor expansion}
g_{\mu\nu}|_{\tau}=g_{\mu\nu}|_{\tau=0}+\sum^\infty_{n=1}\frac{g^{(n)}_{\mu\nu}|_{\tau=0}}{n!}\tau^n,
\end{align}
where the coefficients $g^{(n)}_{\mu\nu}|_{\tau=0}=\left.\frac{\dd^n g_{\mu\nu}}{\dd\tau^n}\right|_{\tau=0}$.
Now, we suppose that each $g^{(n)}_{\mu\nu}$ is given by some combinations of $g_{\mu\nu}|_{\tau}$ and $\till T^{k}_{\mu\nu,\tau}$ ($1\leq k\leq n$), i.e., we give 
\begin{align}
    \label{eq:expansion of gn}g^{(n)}_{\mu\nu}|_{\tau}=c^{(n)}_0g_{\mu\nu}|_{\tau} +\sum^n_{k=1}c^{(n)}_k\till{T}^k_{\mu\nu,\tau},
\end{align}
where coefficients $c^{(n)}_k$ ($0\leq k\leq n$) depend on $\alpha_\tau$ and $\beta_\tau$. 
The condition $g^{(n+1)}_{\mu\nu}=\frac{\dd g^{(n)}_{\mu\nu}}{\dd\tau}$ ensures us to obtain the following recursion relations:
\begin{align}\label{eq:recursion relations}
    \left\{\begin{aligned}
        &c^{(n+1)}_0=\frac{\dd c^{(n)}_0}{\dd \tau}+c^{(n)}_1\beta,\\
        &c^{(n+1)}_k=c^{(n)}_{k-1}+\frac{\dd c^{(n)}_k}{\dd\tau}+c^{(n)}_kk\alpha+c^{(n)}_{k+1}(k+1)\beta,\quad (1\leq k\leq n-1),\\
        &c^{(n+1)}_n=c^{(n)}_{n-1}+c^{(n)}_nn\alpha+\frac{\dd c^{(n)}_n}{\dd\tau},\\
        &c^{(n+1)}_{n+1}=c^{(n)}_n,
    \end{aligned}\right.
\end{align}
where the derivative of $c^{(n)}_k$ ($0\leq k\leq n$) with respect to $\tau$ is given by the chain rule
\begin{align}\label{eq:chain rule}
    \frac{\dd c^{(n)}_k}{\dd \tau}=\frac{\pd c^{(n)}_k}{\pd \alpha}\frac{\dd\alpha}{\dd\tau}+\frac{\pd c^{(n)}_k}{\pd\beta}\frac{\dd\beta}{\dd\tau}.
\end{align}
From Eq.~\eqref{eq:tr[T^m] flow}, the initial values for $c^{(n)}_k$ ($0\leq k\leq n$) are simply given by
\begin{align}
        &c^{(1)}_0=0,&
        &c^{(1)}_1=1,&
        &c^{(2)}_0=\beta,&
        &c^{(2)}_1=\alpha,&
        &c^{(2)}_2=1,
\end{align}
which lead to the solutions for each $g^{(n)}_{\mu\nu}$ as
\begin{align}
    \begin{aligned}
        g^{(n)}_{\mu\nu}|_{\tau=0}=-(-1)^n\left(\frac{2m}{m-1}\right)^{1-n}d^{-1+n}\text{Po}\left(1-\frac{2m}{(m-1)d},-1+n\right)\Big(2m\tr[\bT_0]^{m-1}\Big)^n,
    \end{aligned}
\end{align}
where $\text{Po}(x,y)$ is the Pochhammer symbol defined as $\text{Po}(x,y)=\frac{\Gamma(x+y)}{\Gamma(x)}$. Summing out all $g^{(n)}_{\mu\nu}$ in the expansion of Eq.~\eqref{eq:g Taylor expansion}, we obtain
\begin{align}
    g_{\mu\nu}|_{\tau}
        =&\Big(1+\frac{(m-1)d}{2m}~2 m \tr[\bT_0]^{m-1}\tau\Big)^{\frac{2m}{(m-1)d}}g_{\mu\nu}|_{\tau=0},
\end{align}
which is the same as Eq.~\eqref{trTm g closed form} with $a_m=1$.

To summarize, the flow equations~\eqref{TT system} are the universal forms in any deformation case. In special cases such as $\cO_\tau(T_\tau)=T\bar T$ and $\cO_\tau(T_\tau)=a_m\tr[\bT_\tau]^m$, the flow equation for the metric is directly solvable. Even if the direct solution is not found, we can employ the perturbative method to find the solution $g_{\mu\nu}$ to the flow equation.

Finally, we comment on the perturbative approach to find the solution to the flow equation. In the case of $\cO_\tau(T_\tau)=a_m\tr[\bT_\tau]^m$ discussed above, we have assumed Eq.~\eqref{eq:expansion of gn}. However, this assumption does not always work for other cases.
In general, it is difficult to get the recursion relation like Eq.~\eqref{eq:recursion relations}, so that we need to evaluate $g_{\mu\nu}^{(n)}|_{\tau=0}$ order by order.

\section{Applications of $\mathcal{O}_\tau(T_\tau)=\tr[\bT_\tau]^m$ deformation}
\label{sec: Application of TrT^n}

In this section, we discuss several applications of the deformed theory by $\mathcal{O}_\tau=\tr[\bT_\tau]^m$ in $d$ spacetime dimensions in order to clarify its physical aspects.
In Sec.~\ref{sec:Relations with Gravity}, we argue the relations between the deformed theory by $\mathcal{O}_\tau=\tr[\bT_\tau]^m$ and gravity. We demonstrate that this deformation can be regarded as a transformation from Einstein-Hilbert gravity to a modified gravity. 
In Sec.~\ref{sec:Deformed Lagrangian}, we consider the deformed Lagrangian for a free massless boson in $d$ dimensions and an interactive massive boson in $d=2$.

\subsection{Relation to gravity}\label{sec:Relations with Gravity}

It is well known that the $T\bar{T}$ deformation of the matter field theory in Euclidean $d=2$ is equivalent to the original matter field theory coupled with a gravity theory~\cite{Dubovsky:2017cnj}:
\begin{equation}
\label{eq:TT vs JT}
    S_{T\bar T}=S_{M,\tau_0}[\Psi,\phi]+\int \df^2x \sqrt{g}(\phi R-\Lambda).
\end{equation}
where $S_0[\Psi,\phi]$ is the action for the matter field $\Psi$ and dilaton $\phi$, and $R$ is the Ricci scalar curvature. The second term in the right-hand side of Eq.~\eqref{eq:TT vs JT} is JT-like gravity\footnote{
The action for JT gravity is given by
$$
\int \df^2x \sqrt{g}\phi( R-\Lambda).
$$
} which is a topological gravity defined in two dimensions.

It was shown in Ref.~\cite{Morone:2024ffm} that in higher-dimensional Minkowski spacetimes, an undeformed action $S_{M,\tau_0}$ satisfies the relation
\begin{align}
    \label{eq:graviational equivalence}S_{M,\tau_0}+S_{G,\kappa}\mathop{\simeq}_{g_{\mu\nu}} S_{M,\tau_0+\kappa}+S_\text{EH},
\end{align}
where $S_{M,\tau_0+\kappa}$ is the deformed action with a deformation parameter $\kappa$. Here, $S_{G,\kappa}=\int \df^dx \sqrt{-g} f(R)$ is the $f(R)$-gravity action, while $S_\text{EH}=\int \df^dx\sqrt{-g}R$ for metric $g_{\mu\nu}$ is the Einstein-Hilbert action.
The symbol $\mathop{\simeq}_{g_{\mu\nu}}$ denotes the dynamical equivalence for the metric, i.e., the equations of motion for the metric become the same in both sides in Eq.~\eqref{eq:graviational equivalence}. 
More specifically, the deformed-matter field theory is
\begin{align}
    \label{eq:TT deformation in gravity}S_{M,\tau_0+\kappa}=S_{M,\tau_0}+\int \df^dx\sqrt{-g}\left(f(R)-\frac{\pd f(R)}{\pd R}R\right),
\end{align}
where we have used the fact that through the equation of motion for the metric $g_{\mu\nu}$, the curvature $R$ is related to the stress-energy tensor $T_{\mu\nu,\tau}$ of the matter field $\Psi$ as
\begin{align}\label{eq:E.o.M of T in gravity}
    2\frac{\pd f(R)}{\pd R}R-d\,f(R)=\tr[\bT_0].
\end{align}
Here and hereafter the subscript $0$ in $\bT_0$ means $T_{\mu\nu,\tau}$ at $\tau=\tau_0$.

The authors of Ref.~\cite{Morone:2024ffm} also discussed a specific form of $f(R)$ gravity, namely, Starobinsky gravity in $d=4$ dimensions as follows:
\begin{align}
    \cL_{\text{Star},\kappa}=f(R)=\frac{1}{2}R+\frac{\kappa}{4}R^2,
\end{align}
in which the deformation parameter $\kappa$ provides the mass dimension. 
From Eq.~\eqref{eq:E.o.M of T in gravity}, we obtain 
\begin{align}\label{eq:constrant in Star}
    2\frac{\pd\cL_{\text{Star},\kappa}}{\pd R} R-4\cL_{\text{Star},\kappa}=-R=\tr[\bT_0].
\end{align}
Thus, substituting Eq.~\eqref{eq:constrant in Star} into Eq.~\eqref{eq:TT deformation in gravity}, a deformed matter field theory is given by
\begin{align}\label{eq:Star to TT}
    \begin{aligned}
        S_{M,\tau_0+\kappa}=&S_{M,\tau_0}+\int \df^4x \sqrt{-g}\left(\cL_{\text{Star},\kappa}-\frac{\pd \cL_{\text{Star},\kappa}}{\pd R}R\right)\\
        =&S_{M,\tau_0}-\frac{\kappa}{4}\int \df^4x\sqrt{-g}R^2\\
        =&S_{M,\tau_0}-\frac{\kappa}{4}\int \df^4x\sqrt{-g}\tr[\bT_0]^2.
    \end{aligned}
\end{align}
Here, we set $\kappa=a(\tau-\tau_0)$ with a dimensionless coefficient $a$ to rewrite Eq.~\eqref{eq:Star to TT} as
\begin{align}
    S_{M,\tau}=S_{M,\tau_0}-\frac{a(\tau-\tau_0)}{4}\int \df^4x\sqrt{-g}\tr[\bT_0]^2.
\end{align}
This corresponds just to the $\cO_\tau(T_\tau)=\tr[\bT_\tau]^m$ deformation with $m=2$ which we have discussed in Sec.~\ref{sec:Solutions to flow equation for metric}.

To see the implications more explicitly, let us here consider another $f(R)$ gravity in $d=6$ dimension as
\begin{align}
    f(R)=R+\kappa^2R^3.
\end{align}
The first term is the Einstein-Hilbert term, while the second term can be considered as a higher-order modification. Following the same procedures as in Eq.~\eqref{eq:E.o.M of T in gravity} and Eq.~\eqref{eq:constrant in Star}, we obtain 
\begin{align}\label{eq:E.o.M of trT3 in gravity}
    2\frac{\pd f(R)}{\pd R}R-6f(R)=-4R=\tr[\bT_0],
\end{align}
and then
\begin{align}\label{eq:trT3 in gravity}
    \begin{aligned}
        S_{M,\tau_0+\kappa^2}[g_{\mu\nu},\phi]=&S_{M,\tau}+\int \df^6x\sqrt{-g}\left(f(R)-\frac{\pd f(R)}{\pd R}R\right)\\
        =&S_{M,\tau_0}-2\kappa^2\int \df^6x\sqrt{-g}R^3\\
        =&S_{M,\tau_0}+\frac{\kappa^2}{32}\int \df^6x\sqrt{-g}\tr[\bT_0]^3,
    \end{aligned}
\end{align}
where we have used Eq.~\eqref{eq:E.o.M of trT3 in gravity}. After the replacement $\kappa^2\rightarrow \tau-\tau_0$, we see that
\begin{align}
    S_{M,\tau}=S_{M,\tau_0}+\frac{\tau-\tau_0}{32}\int \df^6x\sqrt{-g}\tr[\bT_0]^3.
\end{align}
Thus, this is the $\cO_\tau(T_\tau)=\tr[\bT_\tau]^m$ deformation with $m=3$ that we have discussed in Sec.~\ref{sec:Solutions to flow equation for metric}.

By following the same calculation as above, we find that the spacetime dimension is related to the power of the deformation operator by $d=2m$. The $f(R)$-gravity theory, given by $f(R)=R+\kappa^{m-1}R^m$, corresponds to a $d=2m$ dimensional deformed field theory triggered by the deformation $\mathcal{O}_\tau(T_\tau)=\text{tr}[\mathbf{T}_\tau]^m$.

From the discussion in this subsection, we conclude that the deformation merely translates the deformation from the gravity side to the field theory side in a composite field theory and a gravity system. If we ignore the contributions from gravity, we return to the scenario discussed in Sec.~\ref{sec:Solutions to flow equation for metric}, which is similar to a quantum field theory in curved spacetime because the metric dynamics are absent.

\subsection{Interactive boson in $d=2$}\label{sec:Deformed Lagrangian}

In this subsection, we deal with the deformed Lagrangian for the $\cO_\tau(T_\tau)=\tr[\bT_\tau]^m$ case and aim to obtain the closed form for the deformed action. 

We consider the interactive massive boson case in $d=2$, and its extension for the potential term.
The undeformed interactive boson Lagrangian is given by
\begin{align}
    \mathcal L_{\tau=0}= g^{\mu\nu}\partial_\mu\phi\partial_\nu\phi-V(\phi),
\end{align}
where we assumed a fixed background metric in the kinetic term.

Now, we consider the deformation operator $\cO_\tau(T_\tau)=\tr[\bT_\tau]^m$ for arbitrary $m$ and show the deformed Lagrangian.
The definition of the stress-energy tensor is given in Eq.~\eqref{def of stress-energy tensor} as
\begin{equation}
    T_{\mu\nu,\tau}
    =-2\frac{\partial \mathcal{L}_\tau}{\partial g^{\mu\nu}} +g_{\mu\nu} \mathcal{L}_\tau,
\end{equation}
from which the trace of the stress-energy tensor reads
\begin{align}
    \tr[\bT_\tau]=-2V_{\tau}(\phi).
\end{align}
Here we denote the deformed interaction by $V_\tau(\phi)$.
Then, from Eq.~\eqref{def of defor}, the deformation for the Lagrangian is expressed by
\begin{align}
    \frac{\df\cL_{\tau}}{\df\tau}=(-2V_\tau(\phi))^m.
    \label{eq:deformed lag}
\end{align}
Note that the kinetic term is invariant under the deformation 
\begin{align}
    \frac{\dd}{\dd\tau}(\partial^\mu\phi\partial_\mu\phi)=0.
\end{align}
Solving Eq.~\eqref{eq:deformed lag}, the deformed Lagrangian is given by
\begin{align}\label{eq:trTm in interactive boson}
    \begin{aligned}
        \cL_\tau=&\partial_\mu\phi\partial^\mu\phi-\frac{V(\phi)}{\Big(1+(-2)^m(m-1)\tau~(V(\phi))^{m-1}\Big)^{1/(m-1)}}.
    \end{aligned}
\end{align}
Note that in the limit $\tau\to\infty$, the Lagrangian \eqref{eq:trTm in interactive boson} reduces to a free boson theory.
This seems to be the same as the result obtained in Ref.~\cite{Morone:2024ffm}; however, in that reference, a deformed metric has been assumed. Therefore, the agreement may be a coincidence.\footnote{
Recently, in Ref.~\cite{Morone:2024sdg}, the deformed action with a deformed metric has been derived as
\begin{align*}
    S_\tau=\int \dd^2x\sqrt{g}\left(g^{\mu\nu}\pd_\mu\phi\pd_\nu\phi-\frac{V(\phi)+ 2^m(m-1)\tau V(\phi)^m}{(1+2^{m}(m-1)\tau V(\phi)^{m-1})^{\frac{m}{m-1}}}\right),
\end{align*}
which is different from our result~\eqref{eq:trTm in interactive boson}.
}

\section{Conclusions}\label{sec:Conclusion}

In this work, we have explored the generalization of $T \bar{T}$ deformations to arbitrary dimensions, focusing on background metric deformations induced by an irrelevant operator at the classical level. A key result of our study is the derivation of a universal flow equation for the background metric under polynomial deformations of the stress-energy tensor, as presented in Eq.~\eqref{metric flow +}. This equation provides a direct connection between the deformation operator and the deformed metric, enabling the evolution of the metric to be determined without requiring auxiliary fields.

We examined specific cases of these deformations, including $\text{tr}[\bT_{\tau}]^3$, $\text{tr}[\bT^3_{\tau}]$, and $\text{tr}[\bT^2_{\tau}]\text{tr}[\bT_{\tau}]^2$, verifying that our results align with those obtained through the auxiliary field method. Additionally, we provided exact solutions to the flow equations in the cases of $T \bar{T}$ deformation and $\text{tr}[\bT_{\tau}]^m$ deformation.

Our findings have several important implications. Notably, the deformation $O_{\tau}(T_{\tau}) = \text{tr}[\bT_{\tau}]^m$ leads to a modified gravity theory of the form $f(R)=R+\kappa^{m-1}R^m$ in $d=2m$ spacetime dimensions. As a concrete example, we demonstrated that the $\text{tr}[T_{\tau}]^3$ deformation corresponds to an $f(R)$ gravity theory with $f(R)=R+\kappa^2 R^3$ in six-dimensional spacetime. Furthermore, we established that this deformation possesses a closed-form Lagrangian when applied to an interacting massive boson system in two dimensions.

Despite these advances, several open questions remain. While we successfully derived the flow equations for general deformations of order $N$, such as $\text{tr}[\bT^3_{\tau}]$ and $\text{tr}[\bT^2_{\tau}]\text{tr}[\bT_{\tau}]^2$, their exact solutions are yet to be determined. Moreover, obtaining a flow equation for more intricate cases, such as the $\sqrt{T \bar{T}}$ deformation, remains an outstanding challenge. Another crucial issue is whether these generalized deformations preserve integrability, a well-known feature of the standard $T \bar{T}$ deformation. Additionally, our analysis is presently limited to the classical level, leaving open the question of how to define these deformed operators in the quantum regime. A promising approach for investigating the quantum aspects of deformed theories is the functional renormalization group framework, which we leave for future study.


\section*{Acknowledgments}

We thank Yue-Peng Guan and Xin-Cheng Mao for their
insightful discussions and valuable contributions to this research project. We also thank Song He, Bum-hoon Lee, Hao Ouyang and Kentaroh Yoshida for their valuable insights and comments.
The work of M.\,Y. is supported by the National Science Foundation of China (NSFC) under Grant No.~12205116 and the Seeds Funding of Jilin University.

\appendix

\section{Auxiliary field method in $T\bar{T}$ deformation within the path integral}
\label{appsec: Auxiliary field}

In Sec.~\ref{sec:Applications of metric flow}, we have introduced an auxiliary field $h_{\mu\nu}$. In this appendix, we show the introduction of the auxiliary field within the path integral formalism.
We start by defining the following 4-vectors
\begin{align}\label{rename}
    \begin{aligned}
        &h_\mu=(h_{11}~h_{12}~h_{21}~h_{22}),\\
        &g^\mu=(g^{11}~g^{12}~g^{21}~g^{22}),\\
        &g_\mu=(g_{11}~g_{12}~g_{21}~g_{22}),\\
        &T^{\mu}_\tau=(T^{11}_\tau~T^{12}_\tau~T^{21}_\tau~T^{22}_\tau),
    \end{aligned}
\end{align}
and matrices
\begin{align}
    G^{\mu\nu}=g^\mu g^\nu,\quad H^{\mu\nu}=\begin{pmatrix}
        g^1g^1 & g^1g^2 & g^2g^1 & g^2g^2\\
        g^1g^3 & g^1g^4 & g^2g^3 & g^2g^4\\
        g^3g^1 & g^3g^2 & g^4g^1 & g^4g^2\\
        g^3g^3 & g^3g^4 & g^4g^3 & g^4g^4
    \end{pmatrix},
\end{align}
and
\begin{align*}
    J^{\mu\nu}=G^{\mu\nu}-H^{\mu\nu}=\begin{pmatrix}
        0 & 0 & 0 & g^1g^4-g^2g^2\\
        0 & g^2g^2-g^1g^4 & 0 & 0\\
        0 & 0 & g^3g^3-g^4g^1 & 0\\
        g^4g^1-g^3g^3 & 0 & 0 & 0
    \end{pmatrix}.
\end{align*}
Thanks to $g^2=g^3$, the matrix $G^{\mu\nu}$ and $H^{\mu\nu}$ are symmetric. 

Using the notations above, the path integral for $h_{\mu\nu}$ with the $T\bar{T}$ action~\eqref{TT auxiliary action} is  
\begin{align}\label{path integral TT}
    \begin{aligned}
        \int \mathcal{D}h_{\mu\nu}e^{-\hatt{S}_{\tau+\delta\tau}[\phi,g_{\mu\nu},h_{\mu\nu}]}
        =&\int\mathcal{D}h_\mu e^{-S_{\tau}[\phi,g_{\mu\nu}]-\delta\tau\int \dd^2x\sqrt{g}\left[ch^\text{T}_\mu(G^{\mu\nu}-H^{\mu\nu})h_\nu-\frac{1}{2}T^{\mu\text{T}}_\tau h_\mu\right]}\\
        =&\int\mathcal{D}h_{\mu}e^{-S_\tau[\phi,g_{\mu\nu}]}\prod_{\dd x}e^{-\delta\tau\sqrt{g}\left[ch^\text{T}_\mu(G^{\mu\nu}-H^{\mu\nu})h_\nu-\frac{1}{2}T^{\mu\text{T}}_\tau h_\mu\right]}.
    \end{aligned}
\end{align}
Because the matrix $G^{\mu\nu}-H^{\mu\nu}$ is symmetric, we can introduce an orthogonal matrix ${\mathcal{O}_\mu}^\nu$ to diagonalize $J^{\mu\nu}$ such that
\begin{align}
    \begin{aligned}
        J^{\mu\nu}=({\mathcal{O}_\rho}^\mu)^{\text{T}}(G^{\rho\sigma}-H^{\rho\sigma}){\mathcal{O}_\sigma}^\nu={\mathcal{O}^\mu}_\rho(G^{\rho\sigma}-H^{\rho\sigma}){\mathcal{O}_\sigma}^\nu,
    \end{aligned}
\end{align}
whose eigenvalues are $J^\mu$. 
Thus, we see that
\begin{align}
\label{eq: pathintegral}
    \begin{aligned}
       \text{Eq.~\eqref{path integral TT}} =&e^{-S_\tau[\phi,g_{\mu\nu}]}\prod_{\dd x}\int\mathcal{D}h_{\mu}\det[\mathcal{O}]e^{-c\delta\tau\sqrt{g}\sum_{\mu}J^\mu h^2_\mu+\frac{1}{2}c\delta\tau\sqrt{g}T^{\nu\text{T}}{\mathcal{O}_\nu}^\mu h_\mu}\\
        =&e^{-S_\tau[\phi,g_{\mu\nu}]}\frac{1}{(2c\delta\tau\sqrt{g})^2}\prod_{\dd x}\int \mathcal{D}h_\mu e^{-\frac{1}{2}\sum_\mu J^\mu h^2_\mu+\sqrt{\frac{c\delta\tau\sqrt{g}}{8}}T^{\nu\text{T}}{\mathcal{O}_\nu}^\mu h_\mu}\\
        =&e^{-S_\tau[\phi,g_{\mu\nu}]}\frac{1}{(2c\delta\tau\sqrt{g})^2}\prod_{\dd x}\left(\frac{(2\pi)^4}{\det[J^{\mu\nu}]}\right)^{1/2}e^{\frac{1}{2}\frac{c\delta\tau\sqrt{g}}{8}\frac{2}{g^1g^4-g^2g^2}(T^1T^4-T^2T^2)}.
    \end{aligned}
\end{align}
Comparing with Eq.~\eqref{rename}, we find that
\begin{align}
    \begin{aligned}
      \text{Eq.~\eqref{eq: pathintegral}}  =&e^{-S_\tau[\phi,g_{\mu\nu}]}\frac{(2\pi)^2}{(2c\delta\tau)^2}\prod_{\dd x}e^{\frac{c}{8}\delta\tau\sqrt{g}(g_{\mu\nu}g_{\rho\sigma}-g_{\mu\rho}g_{\nu\sigma})T^{\mu\nu}_\tau T^{\rho\sigma}_\tau}\\
        =&\text{const}\times e^{-S_\tau[\phi,g_{\mu\nu}]}e^{\frac{c}{8}\delta\tau\int \dd^2x\sqrt{g}(g_{\mu\nu}g_{\rho\sigma}-g_{\mu\rho}g_{\nu\sigma})T^{\mu\nu}_\tau T^{\rho\sigma}_\tau}\\
        =&\text{const}\times e^{-\left(S_\tau[\phi,g_{\mu\nu}]-\frac{c}{8}\delta\tau\int \dd^2x\sqrt{g}(g_{\mu\nu}g_{\rho\sigma}-g_{\mu\rho}g_{\nu\sigma})T^{\mu\nu}_\tau T^{\rho\sigma}_\tau\right)}.
    \end{aligned}
\end{align}
Once we neglect the overall constant, we obtain the $T\bar{T}$-deformed action. Thus, the auxiliary action~\eqref{TT auxiliary action} is an equivalent description of the $T\bar{T}$ deformation.

\section{Auxiliary field method for $\mathcal{O}_\tau(T_\tau)=\tr[\bT_\tau^3]$ and $\mathcal{O}_\tau(T_\tau)=\tr[\bT_\tau^2]\tr[\bT_\tau]^2$}\label{App:T3 and T4}

\subsection{$\mathcal{O}_\tau(T_\tau)=\tr[\bT_\tau^3]$}
For the $\mathcal{O}_\tau(T_\tau)=\tr[\bT_\tau^3]$ deformation, we introduce two auxiliary fields $h_{\mu\nu},E_{\mu\nu}$ to Eq.~\eqref{eq:T3 defor 2nd} such that
\begin{align}
    \begin{aligned}
        &\hatt S_{\tau+\delta\tau}[\phi,g_{\mu\nu},h_{\mu\nu},E_{\mu\nu}]=S_\tau[\phi,g_{\mu\nu}]\\
        &+\delta\tau\int\dd^dx\sqrt{g}\left(a_1 g^{\rho\sigma}h_{\mu\rho}h_{\sigma\nu}T_\tau^{\mu\nu}+a_2 g^{\rho\sigma} g^{\mu\gamma} g^{\nu\delta}h_{\mu\rho}E_{\sigma\nu}E_{\gamma\delta}+ g^{\rho\sigma}h_{\mu\rho}E_{\sigma\nu}T_\tau^{\mu\nu}\right).
    \end{aligned}
\end{align}
This corresponds to the following choice in Eq.~\eqref{eq:Shat}
\begin{align}
        \mathcal{F}[h_{\mu\nu},E_{\mu\nu}]=&a_2 g^{\rho\sigma} g^{\mu\gamma} g^{\nu\delta}h_{\mu\rho}E_{\sigma\nu}E_{\gamma\delta},\\[1ex]
        \mathscr{G}_{\mu\nu}[h_{\mu\nu},E_{\mu\nu}]=&h_{\mu\rho}g^{\rho\sigma}(a_1h_{\sigma\nu}+E_{\sigma\nu}).
\end{align}
Solving the equations of motion for $h_{\mu\nu}$ and $E_{\mu\nu}$, the on-shell values are given respectively by
\begin{align}
    h_{\mu\nu}^*=\frac{1}{8a_1a_2}T_{\mu\nu,\tau},\quad E_{\mu\nu}^*=-\frac{1}{2a_2}T_{\mu\nu,\tau}.
\end{align}
Thus, applying Eqs.~\eqref{eq:dynamical equivalence in g} and \eqref{flow G} we obtain $a_1a_2^2=\frac{1}{64}$. 
We arrive at the flow equation for metric as
\begin{align}
    \frac{\dd g_{\mu\nu}}{\dd\tau}=-6T_{\mu\rho,\tau}g^{\rho\sigma}T_{\sigma\nu,\tau},
\end{align}
which is the same as Eq.~\eqref{eq:T3 flow eq 2nd}.

\subsection{$\mathcal{O}_\tau(T_\tau)=\tr[\bT_\tau^2]\tr[\bT_\tau]^2$}

For the operator $\mathcal{O}_\tau(T_\tau)=\tr[\bT_\tau^2]\tr[\bT_\tau]^2$, we need to introduce four auxiliary fields and the deformed action is rewritten as
\begin{align}
        &\hatt S_{\tau+\delta\tau}[\phi,g_{\mu\nu},L_{\mu\nu},Y_{\mu\nu},X_{\mu\nu},D_{\mu\nu}]=S_\tau[\phi,g_{\mu\nu}]\nonumber\\
        &\quad +\delta\tau\int\dd^dx\sqrt{g}\left(a_1g^{\mu\nu}g^{\rho\sigma}L_{\nu\rho}L_{\mu\sigma}+a_3g^{\mu\nu}g^{\rho\sigma}g^{\gamma\delta}L_{\nu\rho}Y_{\mu\sigma}Y_{\gamma\delta}+g^{\gamma\delta}L_{\mu\nu}Y_{\gamma\delta}T_{\tau}^{\mu\nu}\right.\nonumber\\
        &\quad \left. + a_2g^{\mu\nu}g^{\rho\sigma}X_{\nu\rho}X_{\mu\sigma}+a_4g^{\mu\nu}g^{\rho\sigma}g^{\gamma\delta}X_{\nu\rho}D_{\mu\sigma}D_{\gamma\delta}+g^{\gamma\delta}g^{\rho\sigma}g_{\mu\nu}X_{\delta\rho}D_{\gamma\sigma}T_\tau^{\mu\nu}\right).
\end{align}
This corresponds to the following choice in Eq.~\eqref{eq:Shat}:
\begin{align}
        \mathcal{F}[L_{\mu\nu},Y_{\mu\nu},X_{\mu\nu},D_{\mu\nu}]=&a_1g^{\mu\nu}g^{\rho\sigma}L_{\nu\rho}L_{\mu\sigma}+a_3g^{\mu\nu}g^{\rho\sigma}g^{\gamma\delta}L_{\nu\rho}Y_{\mu\sigma}Y_{\gamma\delta}\nonumber\\
        &+a_2g^{\mu\nu}g^{\rho\sigma}X_{\nu\rho}X_{\mu\sigma}+a_4g^{\mu\nu}g^{\rho\sigma}g^{\gamma\delta}X_{\nu\rho}D_{\mu\sigma}D_{\gamma\delta},\\[1ex]
        \mathscr{G}_{\mu\nu}[L_{\mu\nu},Y_{\mu\nu},X_{\mu\nu},D_{\mu\nu}]=&g^{\gamma\delta}L_{\mu\nu}Y_{\gamma\delta}+g^{\gamma\delta}g^{\rho\sigma}g_{\mu\nu}X_{\delta\rho}D_{\gamma\sigma}.
\end{align}
Solving the equation of motions~\eqref{eq:EOM for auxiliary fields} for $L_{\mu\nu},Y_{\mu\nu},X_{\mu\nu}$ and $D_{\mu\nu}$, we obtain the on-shell values as
\begin{align}\label{eq:on-shell values T4}
    L_{\mu\nu}^*=\frac{\tr[\bT_\tau]}{8a_1a_3}T_{\mu\nu,\tau},\quad Y_{\mu\nu}^*=-\frac{1}{2a_3}T_{\mu\nu,\tau},\quad X_{\mu\nu}^*=\frac{\tr[\bT_\tau]}{8a_2a_4}T_{\mu\nu,\tau},\quad D_{\mu\nu}^*=-\frac{1}{2a_4}T_{\mu\nu,\tau}.
\end{align}
Substituting Eq.~\eqref{eq:on-shell values T4} to Eqs.~\eqref{eq:dynamical equivalence in g} and \eqref{flow G}, we find $a_1a_3^2=a_2a_4^2=-\frac{1}{32}$ and obtain the flow equation for the metric
\begin{align}
    \frac{\dd g_{\mu\nu}}{\dd\tau}=-4(\tr[\bT_\tau]^2T_{\mu\nu,\tau}+\tr[\bT_\tau^2]\tr[\bT_\tau]g_{\mu\nu}),
\end{align}
where $\tr[\bT_\tau^2]=g^{\mu\nu}g^{\rho\sigma}T_{\mu\rho,\tau}T_{\sigma\nu,\tau}$. This flow equation is the same as Eq.~\eqref{eq:flow eq T4}.

\section{Some derivations}\label{appsec: detail of derivations}

In this appendix, we provide detailed derivations of the equations presented in the main text.

\subsection*{Eq.~\eqref{eq: derivative T with g}}

From the definition of stress-energy tensor~\eqref{def of stress-energy tensor}
\begin{align}
        \frac{\pd T^{\rho\sigma}_\tau}{\pd g_{\mu\nu}}=&\frac{\pd}{\pd g_{\mu\nu}}\left(-\frac{2}{\sqrt{g}}\frac{\delta S_\tau[\phi,g_{\mu\nu}]}{\delta g_{\rho\sigma}}\right)\nonumber\\
        =&\frac{1}{\sqrt{g}}g^{\mu\nu}\frac{\delta S_\tau[\phi,g_{\mu\nu}]}{\delta g_{\rho\sigma}}-\frac{2}{\sqrt{g}}\frac{\pd}{\pd g_{\rho\sigma}}\frac{\delta S_\tau[\phi,g_{\mu\nu}]}{\delta g_{\mu\nu}}\nonumber\\
        =&-\frac{1}{2}g^{\mu\nu}T^{\rho\sigma}_\tau-\frac{2}{\sqrt{g}}\frac{\pd}{\pd g_{\rho\sigma}}\left(-\frac{\sqrt{g}}{2}T^{\mu\nu}_\tau\right)\nonumber\\
        =&-\frac{1}{2}g^{\mu\nu}T^{\rho\sigma}_\tau+\frac{1}{2}g^{\rho\sigma}T^{\mu\nu}_\tau+\frac{\pd T^{\mu\nu}_\tau}{\pd g_{\rho\sigma}},
\end{align}
where we have used
\begin{align}\label{eq:var of g}
    \delta g=gg^{\mu\nu}\delta g_{\mu\nu},
\end{align}
and 
\begin{align}
\frac{\pd}{\pd g_{\mu\nu}}\frac{\delta S_\tau[\phi,g_{\mu\nu}]}{\delta g_{\rho\sigma}}=\frac{\pd}{\pd g_{\rho\sigma}}\frac{\delta S_\tau[\phi,g_{\mu\nu}]}{\delta g_{\mu\nu}}.
\end{align}

\subsection*{Eq.~\eqref{sqrt metric}}

Using Eqs.~\eqref{eq:var of g} and \eqref{TT system} the first order derivative for the determinant of metric is given by
\begin{align}
\frac{\dd\sqrt{g}}{\dd\tau}=\sqrt{g}\,\tr[\bT_\tau].
\end{align}
Since $\frac{\dd}{\dd\tau}(g^{\mu\nu}g_{\nu\rho})=0$, the flow equation for inverse metric $g^{\mu\nu}$ reads
\begin{align}
    \frac{\dd g^{\mu\nu}}{\dd\tau}=-2\tr[\bT_\tau]g^{\mu\nu}+2T^{\mu\nu}_\tau.
\end{align}
Thus the second order derivative for the determinant of the metric is given by
\begin{align}
    \frac{\dd^2\sqrt{g}}{\dd\tau^2}=\frac{\dd}{\dd\tau}(\sqrt{g}g^{\mu\nu}T_{\mu\nu,\tau})=\sqrt{g}(\tr[\bT_\tau]^2-\tr[\bT^2_\tau])=\sqrt{g}\mathcal{O}_\tau(T_\tau).
\end{align}
The third order derivative is
\begin{align}
        \frac{\dd^3\sqrt{g}}{\dd\tau^3}=&\frac{\dd}{\dd\tau}(\sqrt{g}\mathcal{O}_\tau(T_\tau))
        \nonumber\\
        =&\sqrt{g}\tr[\bT_\tau]\mathcal{O}_\tau(T_\tau)+\sqrt{g}\left(2\tr[\bT_\tau]\frac{\dd}{\dd\tau}(g^{\mu\nu}T_{\mu\nu,\tau})-\frac{\dd}{\dd\tau}(g^{\mu\nu}T_{\nu\rho,\tau}g^{\rho\sigma}T_{\sigma\mu,\tau})\right)=0.
\end{align}

\subsection*{Eq.~\eqref{TrT}}

The derivative of $\tr[\bT_\tau]$ in $\mathcal{O}_\tau(T_\tau)$ deformation comes from Eq.~\eqref{eq:flow equation for g and T} directly
\begin{align}
        \frac{\dd\tr[\bT_\tau]}{\dd\tau}=&\frac{\dd}{\dd\tau}(g_{\mu\nu}T^{\mu\nu,\tau})\nonumber\\
        =&2T^{\mu\nu}_\tau\frac{\pd\mathcal{O}_\tau(T_\tau)}{\pd T^{\mu\nu}_\tau}+g_{\mu\nu}\left[(g^{\mu\nu}T^{\rho\sigma}_\tau-g^{\rho\sigma}T^{\mu\nu}_\tau)\frac{\pd\mathcal{O}_\tau(T_\tau)}{\pd T^{\rho\sigma}_\tau}-g^{\mu\nu}\mathcal{O}_\tau(T_\tau)-\frac{\pd\mathcal{O}_\tau(T_\tau)}{\pd g_{\mu\nu}}\right] \nonumber\\
        =&(dT^{\mu\nu}_\tau-\tr[\bT_\tau]g^{\mu\nu})\frac{\pd\mathcal{O}_\tau(T_\tau)}{\pd T^{\mu\nu}_\tau}-d\mathcal{O}_\tau(T_\tau),
\end{align}
where we have used the fact that for homogeneous $\mathcal{O}_\tau(T_\tau)$,
\begin{align}
    g_{\mu\nu}\frac{\pd\mathcal{O}_\tau(T_\tau)}{\pd g_{\mu\nu}}=T^{\mu\nu}_\tau\frac{\pd\mathcal{O}_\tau(T_\tau)}{\pd T^{\mu\nu}_\tau}.
\end{align}

\bibliographystyle{JHEP}
\bibliography{ref}

\end{document}